\documentclass{webofc}
\usepackage[varg]{txfonts}   
\begin{document}
\title{The Interaction of Neutrons With $^7$Be: Lack of Standard
Nuclear Physics  Solution to the ``Primordial $^7$Li Problem"}
%
%
\subtitle{On Behalf of the SARAF Israel-US-Switzerland Collaboration \\ (Listed in Ref. \cite{NPA8})}

\author{\firstname{Moshe} \lastname{Gai}\inst{1}\fnsep\thanks{\email{moshe.gai@uconn.edu}}
}
\institute{LNS at Avery Point, University of Connecticut, 1084 Shennecossett Rd., Groton, CT 06340, USA.}

\abstract{The destruction of $^7$Be with neutrons represents the last possible standard avenue to reduce the predicted abundance of the primordial $^7$Li and in this way to attempt to solve the Cosmological $^7$Li problem. We discuss the results of an experiment performed at the Soreq Applied Research Accelerator Facility (SARAF) in Israel where we measured the Maxwellian Averaged Cross Sections (MACS) of the $^7$Be(n,p),  $^7$Be(n,$\alpha$), and $^7$Be(n,$\gamma \alpha)$ reactions. Our MACS measured at 49.5 keV in the window of the Big Bang Nucleosynthesis (BBN), indicate the lack of standard nuclear physics solution to the ``Primordial $^7$Li Problem".}
\maketitle
\section{Introduction}
\label{intro}

Standard cosmology predicts the original (primordial) chemical composition of the Universe very precisely, as the result of a brief period of Big Bang Nucleosynthesis (BBN) \cite{BBN1,BBN2,BBN3}.  In a minimal model (only standard model particles, no large lepton/antilepton asymmetry), the only astrophysical input to the BBN calculation is the baryon density of the Universe, which is now known precisely \cite{WMAP}. Within plausible errors, the observed abundances of helium and especially deuterium are in good agreement with the BBN predictions at the known baryon density.  The only other sources of significant uncertainty in the standard model of BBN are the cross sections of the twelve ``canonical" BBN nuclear reactions \cite{NUCL}. With total errors of the order a few percent or less, the BBN predictions are a very specific consequence of modern cosmology and in that sense BBN is one of the most remarkable achievements of modern cosmology and nuclear astrophysics.

However, early on it was already noticed that BBN theory fails to predict correctly the observed abundance of $^7$Li. It over predicts the abundance of $^7$Li by approximately a factor of three and up to five sigma deviation from observation  \cite{7Li}. This disagreement is very difficult to understand in terms of cosmology and it has been dubbed the ``Primordial $^7$Li Problem". 

Approximately 95\% of the primordial $^7$Li  is the byproduct of the electron capture beta-decay of the primordial $^7$Be that occurred approximately a hundred years after its formation when the plasma cooled down enough for the $^7$Be to capture an electron. Hence in a search for a standard nuclear physics solution to the $^7$Li problem great deal of attention was given to measuring the direct destruction of $^7$Be by deuterium \cite{Angulo} or $^3$He \cite{3He} that are also present during BBN. These studies did not reveal cross sections sufficiently large to compete with the standard indirect destruction of $^7$Be in the reaction chain $^7$Be(n,p)$^7$Li(p,$\alpha$) that is included in BBN. Additional searches for new resonances in the d + $^7$Be system \cite{d7Be1,d7Be2} did not yield sufficient destruction of $^7$Be to solve the ``Primordial $^7$Li Problem".

One avenue (the last one that is still not ruled out) for the direct destruction of $^7$Be is by neutrons that are also present during BBN. This possibility has been ignored due to the very small rate compiled by Wagoner in 1969 \cite{Wag69}. For example turning off Wagoner rate for the $^7$Be(n,$\alpha$) reaction in BBN  calculations changes the $^7$Li abundance by only 1\% due to the small direct destruction of $^7$Be by neutrons. 

However Serpico {\em et al.} \cite{Serp} eloquently stated: ``...we adopted Wagoner's rate \cite{Wag69}, assuming a factor of ten uncertainty, as he suggested as a typical conservative value. Within this allowed range, [the interactions of neutrons with $^7$Be] could play a non-negligible role in direct $^7$Be destruction, so it would be fruitful to have a new experimental determination. Apart from the role of unknown or little known $^8$Be resonances, it is however unlucky that the used extrapolation may underestimate the rate by more than one order of magnitude, as this process mainly proceeds through a p-wave". Indeed the possibility of a hitherto unknown resonance in the BBN window (of $\sim$ 40 - 70 keV) at high excitations ($\sim$19.5 MeV) in $^8$Be that would lead to large cross section for the direct destruction of $^7$Be and a standard nuclear physics solution to the ``Primordial $^7$Li Problem" has not been ruled out as of yet. 

\section{The SARAF}
\label{sec1}

The Soreq Applied Research Accelerator Facility (SARAF) \cite{SARAF}, at the Soreq Nuclear Research Center in Israel, offers a major research opportunity for studies of neutron interactions that are important for Stellar Evolution theory and Cosmology. An important advantage of the SARAF (in phase I) is that the neutron beam produced by the Liquid Lithium Target (LiLiT) \cite{LiLiT,LiLiT2,Halfon} via the thick-target $^7$Li(p,n)$^7$Be reaction near threshold, has a quasi-Maxwellian energy distribution peaked at tens keV \cite{Max1,Max2}, and thus mimics cosmological and stellar conditions. The neutrons produced at SARAF with an ``effective temperature" kT = 49.5 keV (energies spread from 1 to  180 keV) are ideally suited for example to measure neutron interactions with $^7$Be in the Big Bang Nucleosynthesis (BBN) window of T = 0.5 - 0.8 GK and kT = 43 - 72 keV. Measurements of neutron interaction with $^7$Be in the BBN window \cite{7Be} are essential for shedding light on the ``Primordial $^7$Li Problem" in order to estimate the rate of destruction of $^7$Be during BBN. This provides strong motivation for the development of experimental protocols that enable the use of robust detectors such as polymer based Nuclear Track Detectors (NTD). 

The resulting high neutron flux, as large as $\sim 10^{10}$ n/sec/cm$^2$ and the associated large flux of 477 keV gamma-rays from the $^7$Li(p,p'$\gamma$) reaction ($\sim 10^{11}\  \gamma$/s), and of 14.6, 17.6 MeV gamma-rays from the $^7$Li(p,$\gamma$) reaction ($\sim10^9 \ \gamma$/s), present a major challenge for detector systems, even for detectors  that can survive in such a ``hostile" neutron and gamma-ray environment. In this first study we use the plastic polymer commonly known as CR39, Columbia Resin \#39, poly allyl diglycol carbonate - PADC, C$_{12}$H$_{18}$O$_7$, Nuclear Track Detectors (NTD) purchased from Homalite$^{TM}$. In addition to neutron background the thermalized neutron capture on the surrounding material yield a large flux of ``environmental" gamma-background. Since D$\Omega /4 \pi \ \sim \ 1.5 \times 10^{-3}$ for the CR39, the ``environmental" gamma-background is the leading source of gamma-background.  

\subsection{The SARAF Measurement}
\label{sec2}

The intense background from neutrons and gamma-rays excluded our attempt for measurements using standard spectroscopic tools such as silicon detectors. We are currently pursuing the use of diamond detectors \cite{Diamond} that are known to be robust against the large neutron and gamma ray flux at SARAF. However, diamond detectors were not used in the current measurement. At present, while the diamond detector system is under development, we used  CR39 plates-NTD, for detecting the alpha-particles and protons from the interaction of the SARAF/LiLiT intense neutron beams with a $^7$Be target, as shown in Figure 1.

\subsection{The SARAF Results and Accomplishments}

The setup of the SARAF measurement is shown in Figure 1 and complete description of the SARAF measurement is in the process of being formulated for publications and we only list here our results and accomplishments:

\begin{figure*}
\centering
\sidecaption
\includegraphics[width=6cm,clip]{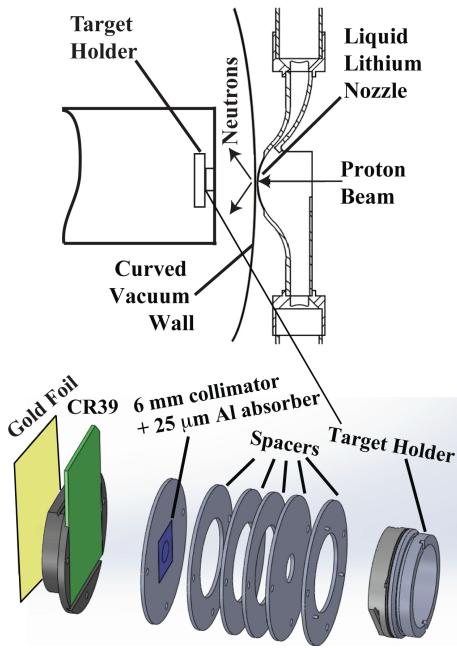}
\caption{Schematic diagram of the experimental setup used at SARAF including the secondary Be target holder and the primary LiLiT target \cite{LiLiT,LiLiT2}.}
\label{fig1}       
\end{figure*}

\itemize

\item{1.}  The etched CR39 plates were calibrated using (3.18 MeV) alpha-particles from a $^{148}$Gd radioactive source and RBS of 1.4 MeV alpha-particles and 1.5 MeV protons.

\item{2.}  The radii region of interest (RRI) of pits observed in the etched CR39 plates that were produced by 1.5 MeV  protons (0.8 -  1.4 $\mu$m) and 1.4 - 3.18 MeV alpha-particle (1.4 - 3.4 $\mu$m) was established.

\item{3.}  A measurement with cold neutrons at the ILL established the background in the proton RRI to be due to gamma-rays and in the alpha-particle RRI the background is due to $\alpha \ + \ ^{14}$C tracks from the $^{17}$O(n,$\alpha)^{14}$C reaction inside the CR39 NTD.

\item{4.}  The MACS of the $^{10}$B(n,$\alpha$) reaction at 49.5 keV was measured to be 3.3$\pm$0.3 b, in perfect agreement with the known MACS vale of 3.35 b.

\item{5.}  A measurement with a $^9$Be target established the background for the measurement with the $^7$Be target.

\item{6.}  The spectrum of radii measured for alpha-particles at SARAF with 49.5 keV neutron beams is in good agreement with the scaled alpha-particle spectrum measured with cold neutrons (with radius scaled by $\sqrt 2$ and the  yield scaled by 13.4). 

\item{7.}  A 25 micron aluminum foil was used to stop the 1.5 MeV protons and measure the yield of the high energy alpha-particles: 9.5 MeV and $\sim$8.4 MeV (that are degraded to 1.0 - 3.3 MeV) from the $^7$Be(n,$\alpha$) and the $^7$Be(n,$\gamma \alpha$) reactions, respectively. The stopper foil also stopped the 1.5 MeV alpha-particles.

\item{8.} No excess counts beyond the background measured with the $^9$Be target was observed for the high energy alpha-particles with energies of 9.5 MeV and $\sim$8.4 MeV.

\item{9.} We measured an upper limit of 0.9 mb for the cross section of the $^7$Be(n,$\alpha$) reaction and the $^7$Be(n,$\gamma_{3,4}$) reactions.

\item{10.} The measured upper limit of 0.9 mb is in perfect agreement with the upper limit measured by the n\_TOF collaboration at 12.7 keV \cite{nTOF}, but it is a factor of 7.4 below the extrapolated value of Hou {\em et al.} \cite{Hou} and the Kyoto measurement \cite{Kyoto}. 

\item{11.} A comparison of concurrent measurements that were performed simultaneously with the absorber foil in place and without the absorber foil (foil in/ foil out measurement), allowed us to measure the 1.5 MeV protons from the $^7$Be(n,p) reaction and the 1.5 MeV alpha-particles from the $^7$Be(n,$\gamma_1)^8$Be$^*$(3.03 MeV) reaction. 

\item{12.} We measure the MACS of the $^7$Be(n,p) reaction at 49.5 keV to be 10.1$\pm 1.0$ (stat), $\pm 1.5$ (syst) in perfect agreement with the extrapolated cross section measured up to 13.5 keV \cite{7Benp}.

\item{13.} We measure the MACS of the $^7$Be(n,$\gamma_1)^8$Be$^*$(3.03 MeV) to be 16.4$\pm 6$ mb, which is a factor of 11.7 larger than the extrapolated value of the n\_TOF collaboration \cite{nTOF}.

\item{14.} From the measured cross section of the $^7$Be(n,$\gamma_1)^8$Be$^*$(3.03 MeV) reaction we extract the B(E1:~$2^- \rightarrow 2^+_1$) = 0.040(14) W.u. which is in excellent accordance with the measured values of the B(E1:~$2^- \rightarrow 2^+_2$) = 0.053 W.u. and B(E1:~$2^- \rightarrow 2^+_3$) = 0.047 W.u. \cite{8Be}. 

\item{15.} Our measurement indicate s-wave dominance in the BBN window in contrast to the Kyoto and n\_TOF measurements that together indicate a p-wave dominance in the BBN window.

\item{16.} Based on all currently available data (and based only on data) we evaluate new $^7$Be(n,$\alpha$) burning rate. Using Wagoner notation \cite{Wag69} we derive $g_1 \ = \ 112 \ + \ 34$T (T in GK). We add to this ``new Wagoner-like" rate the $2^+$ resonance at 20.1 MeV in $^8$Be, as shown in Figure 2. This rate is almost a factor of 10 smaller then the original Wagoner rate \cite{Wag69} currently used in all BBN calculations. 

\begin{figure}
\centering
\sidecaption
\includegraphics[width=10cm,clip]{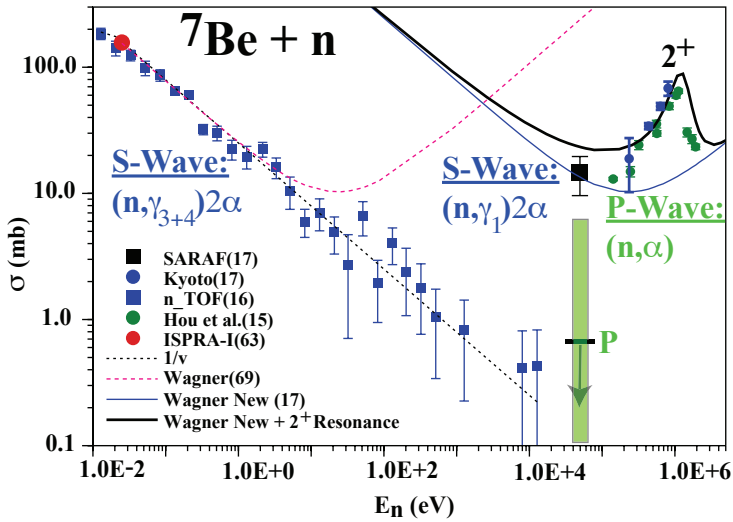}
\caption{Current world data \cite{7Be,nTOF,Hou,Kyoto,Ispra} on alpha-particles emanating from the interaction of neutrons with $^7$Be compared to Wagoner's MACS (long dashed purple line) \cite{Wag69}. The New Wagoner-like MACS required by our data (shown in Blue) plus the 2$^+$ resonance is shown (in black) as discussed in the text. The Big Bang window for the formation of $^7$Be \cite{NUCL} is indicated by a (light green) vertical full rectangle and the SARAF result \cite{7Be} is indicated in black square.}
\label{fig2}       
\end{figure}

\section{Conclusions}

We measured the MACS of the $^7$Be(n,$\alpha$), $^7$Be(n,$\gamma \alpha$) and $^7$Be(n,p) reactions in BBN window. We report the first Measurement of $^7$Be(n,$\alpha)^8$Be$^*$(3.03 MeV) reaction from which we extract  B(E1: $2^- \rightarrow 2^+_1$) = 0.04 W.u. Our measured MACS value of the $^7$Be(n,p) reaction at 49.5 keV of 10.1 $\pm 1.0$ (stat) $\pm 1.5$ (syst) b is in good agreement with the extrapolated value measured up to 13.5 keV. Our measured MACS in the BBN window does not agree with previous s and p waves extrapolations into the BBN window and we conclude neutron s - wave dominance in the BBN window. We provide new Wagoner-like burning rate of the $^7$Be(n,$\alpha$) reaction which is based on world data solely \cite{7Be,nTOF,Hou,Kyoto,Ispra}. We did not observe a hitherto unknown resonance in the BBN window that would lead to large cross section, large enough to solve the ``Primordial $^7$Li Problem" and we conclude on lack of standard nuclear solution of the ``Primordial $^7$Li Problem".

\section{ Post Script Remark}
The current paper presents the results of the SARAF US-Israel-Switzerland collaboration that were approved for public presentation in the DNP meeting of the American Physical Society in Vancouver, Canada on October 14, 2016 \cite{7Be}. All members of the collaboration listed as co-authors in \cite{7Be} approved the results presented in the DNP meeting in October 2016. In addition this invited talk by the author and a second poster paper by E.E. Kading {\em et al.} in this NPA8 meeting \cite{NPA8}, was approved by the collaboration on February 2, 2017. Specifically all co-authors listed the NPA8 poster paper \cite{NPA8} approved on February 4, 2017 the paper \cite{NPA8} which also referenced the invited talk of the author. On March 14, 2017, the collaboration learned that two colleagues had a change of mind. Their claims were seriously considered by the collaboration, some changes were adopted, but the essential claims made in the communication of March 14, 2017, were refuted by the rest of the collaboration on March 24, 2017. While the discussion is still going on, the collaboration did not as of yet received convincing argument that invalidates the material that was approved for public announcement already a year ago in October 2016 and again for the second time in this NPA8 meeting. The collaboration is committed to continue the internal scientific dialogue and we will continue to judge statements of facts based on their scientific merit.





\section{Acknowledgement}
The material presented in this paper is based upon work supported by the U.S.-Israel Bi National Science Foundation, Award Number 2012098, and the U.S. Department of Energy, Office of Science, Office of Nuclear Physics, Award Numbers: DE-FG02-94ER40870 and DE-FG02-91ER-40608.
%
%
%


\end{document}